\documentclass{amsart}
\usepackage{amsfonts}

\newtheorem{theorem}{Theorem}
\theoremstyle{plain}

\numberwithin{equation}{section}

\input{tcilatex}

\begin{document}
\title{On Stochastic Generators of Completely Positive Cocycles.}
\author{V. P. Belavkin.}
\address{Mathematics Department, University of Nottingham, \\
NG7 2RD, UK.}
\email{vpb@maths.nott.ac.uk}
\date{July 20, 1994}
\subjclass{Stochastic Analysis}
\keywords{Quantum Stochastic Generators, Completely Positive Cocycles,
Quantum Filtering Equations.}
\thanks{Published in: \textit{Russian Journal of Math. Phys}. \textbf{3}
(1995) No 4, 523--528}
\maketitle

\begin{abstract}
A characterisation of the generators of quantum stochastic cocycles of
completely positive (CP) maps is given in terms of the complete
dissipativity (CD) of its form-generator. The pseudo-Hilbert dilation of the
stochastic form-generator and the pre-Hilbert dilation of the corresponding
dissipator is found. The general form of the linear continuous structural
maps for the algebra of all bounded operators is derived and the quantum
stochastic flow for the corresponding cocycle is outlined. It is proved that
any w*-analytical bounded CD form-generator give rise to a quantum
stochastic CP cocycle over a von Neumann algebra.
\end{abstract}

\section{Quantum Stochastic CP Cocycles and their Generators.}

The quantum filtering theory \cite{1} provides examples for a new type of
irreversible quantum dynamics, described by one-parameter cocycles: $\phi
=\left( \phi _t\right) _{t>0}$ of completely positive stochastic maps $\phi
_t\left( \omega \right) :\mathcal{B}\rightarrow \mathcal{B}$ of an operator
algebra $\mathcal{B}\subseteq \mathcal{B}\left( \mathfrak{h}\right) $. The
cocycle condition 
\begin{equation*}
\phi _s\left( \omega \right) \circ \phi _r\left( \omega ^s\right) =\phi
_{r+s}\left( \omega \right)
\end{equation*}
means the stationarity, with respect to the shift $\omega ^s=\left\{ \omega
\left( t+s\right) \right\} $ of a given stochastic process $\omega =\left\{
\omega \left( t\right) \right\} $. Such maps are in general unbounded, but
normalized, $\phi _t\left( I\right) =M_t$ to an operator-valued martingale $%
M_t=\epsilon _t\left[ M_s\right] \geq 0$ with $M_0=1$, or a positive
submartingale: $M_t\geq \epsilon _t\left[ M_s\right] $, for all $s>t$, where 
$\epsilon _t$ is the conditional expectation with respect to the history up
to time $t$.

In the most general case, the stochastically differentiable family $\phi $
with respect to a quantum stationary process, with independent increments $%
A^s\left( t\right) =A\left( t+s\right) -A\left( s\right) $ generated by a
finite dimensional It\^{o} algebra is described by the quantum stochastic
equation 
\begin{equation*}
\mathrm{d}\phi _t\left( X\right) =\phi _t\circ \alpha _\nu ^\mu \left(
X\right) \mathrm{d}A_\mu ^\nu :=\sum_{\mu ,\nu }\phi _t\left( \alpha _\nu
^\mu \left( X\right) \right) \mathrm{d}A_\mu ^\nu ,\qquad \text{ }X\in 
\mathcal{B}\qquad (1)
\end{equation*}
with the initial condition $\phi _0\left( X\right) =X,$ for all $X\in 
\mathcal{B}$. Here $A_\mu ^\nu \left( t\right) $ with $\mu \in \left\{
-,1,...,d\right\} ,\quad $ $\nu \in \left\{ +,1,...,d\right\} $ are the
standard time $A_{-}^{+}\left( t\right) =tI$, annihilation $A_{-}^m\left(
t\right) $, creation $A_n^{+}\left( t\right) $ and exchange $A_n^m\left(
t\right) $ operator integrators with $m,n\in \left\{ 1,...,d\right\} $. The
infinitesimal increments $\mathrm{d}A_\nu ^\mu \left( t\right) =A_s^{t\mu
}\left( \mathrm{d}t\right) $ are formally defined by the
Hudson-Parthasarathy multiplication table \cite{2} and the $\flat $-property 
\cite{3}, 
\begin{equation*}
\mathrm{d}A_\mu ^\beta \mathrm{d}A_\gamma ^\nu =\delta _\gamma ^\beta 
\mathrm{d}A_\mu ^\nu ,\qquad \text{ }A^{\flat }=A,\qquad \left( 2\right)
\end{equation*}
where $\delta _\gamma ^\beta $ is the usual Kronecker delta over the indices 
$\beta \in \left\{ -,1,...,d\right\} ,\quad \gamma \in \left\{
+,1,...,d\right\} $and $A_{-\nu }^{\flat \mu }=A_{-\mu }^{\nu \dagger }$
with respect to the reflection $-(-)=+,$ $-(+)=-$ of the indices $\left(
-,+\right) $ only, such that $-m=m,\quad -n=n$. The structural maps $\alpha
_\nu ^\mu :\mathcal{B}\rightarrow \mathcal{B}$ for the $*$-cocycles $\phi
_t^{*}=\phi _t$, where $\phi _t^{*}\left( X\right) =\phi _t\left( X^{\dagger
}\right) ^{\dagger }$ should obviously satisfy the $\flat $-property $\alpha
^{\flat }=\alpha $, where $\alpha _{-\mu }^{\flat \nu }=\alpha _{-\nu }^{\mu
*},$ $\alpha _\nu ^{\mu *}\left( X\right) =\alpha _\nu ^\mu \left(
X^{\dagger }\right) ^{\dagger }$, even in the case of nonlinear $\phi _t$
and so $\alpha _\nu ^\mu $. If the coefficients $\alpha _\nu ^\mu $ are
independent of $t$, $\phi $ satisfies the cocycle property $\phi _s\circ
\phi _r^s=\phi _{s+r},$ where $\phi _t^s$ is the solution to (1) with $A_\nu
^\mu \left( t\right) $ replaced by $A_\nu ^{s\mu }\left( t\right) .$

Let us prove that the ''spatial'' part $\boldsymbol{\lambda }=\left( \lambda
_\nu ^\mu \right) _{\nu \neq -}^{\mu \neq +}$ of the tensor $\lambda =\alpha
+\delta $ for a CP cocycle $\phi $ with $\delta _\nu ^\mu \left( X\right)
=X\delta _\nu ^\mu $ must be conditionally CP in the following sense.

\begin{theorem}
Suppose that the quantum stochastic equation (1) with $\phi _0\left(
X\right) =X$ has a completely positive solution in the sense of positive
definiteness of the matrix $\left[ \phi _t\left( X_{kl}\right) \right]
,\quad \forall t>0$ given by an arbitrary positive definite matrix $\left[
X_{kl}\right] $ with the elements $X_{kl}\in \mathcal{B}$. Then the matrix $%
\boldsymbol{\lambda }$ of structural maps $\lambda _\nu ^\mu \left( X\right)
=\alpha _\nu ^\mu \left( X\right) +\delta _\nu ^\mu \left( X\right) $ is
conditionally completely positive, 
\begin{equation*}
\sum_{k,l}\langle \boldsymbol{\eta }_k|\boldsymbol{\iota }\left(
X_{kl}\right) \boldsymbol{\eta }_l\rangle =0\Longrightarrow
\sum_{k,l}\langle \boldsymbol{\eta }_k|\boldsymbol{\lambda }\left(
X_{kl}\right) \boldsymbol{\eta }_l\rangle \geq 0
\end{equation*}
where $\left[ X_{kl}\right] \geq 0,\boldsymbol{\eta }\in \mathfrak{h}\otimes 
\mathbb{C}^{d+1}$ with respect to the degenerate representation $\boldsymbol{%
\iota }=\left( \iota _\nu ^\mu \right) _{\nu \neq -}^{\mu \neq +},$ $\iota
_\nu ^\mu \left( X\right) =X\delta _\nu ^{+}\delta _{-}^\mu $, both written
in the matrix form as 
\begin{equation*}
\boldsymbol{\lambda }=\left( 
\begin{array}{cc}
\lambda & \lambda _{\bullet } \\ 
\lambda ^{\bullet } & \lambda _{\bullet }^{\bullet }%
\end{array}
\right) ,\qquad \text{ }\boldsymbol{\iota }\left( X\right) =\left( 
\begin{array}{cc}
X & 0 \\ 
0 & 0%
\end{array}
\right) \qquad \left( 3\right)
\end{equation*}
with $\lambda =\alpha _{+}^{-},\quad $ $\lambda ^m=\alpha _{+}^m,\quad $ $%
\lambda _n=\alpha _n^{-},\quad \lambda _n^m=\delta _n^m+\alpha _n^m,$where $%
\delta _n^m\left( X\right) =X\delta _n^m,$ such that 
\begin{equation*}
\lambda \left( X^{\dagger }\right) =\lambda \left( X\right) ^{\dagger
},\qquad \text{ }\lambda ^n\left( X^{\dagger }\right) =\lambda _n\left(
X\right) ^{\dagger },\qquad \text{ }\lambda _n^m\left( X^{\dagger }\right)
=\lambda _m^n\left( X\right) ^{\dagger }.\left( 4\right)
\end{equation*}
\proof
Let us denote by $\mathcal{D}$ the $\mathfrak{h}$-span $\left\{ \sum_f\xi
^f\otimes f^{\otimes }\left| \xi ^f\in \mathfrak{h},f^{\bullet }\in \mathbb{C%
}^d\otimes L^2\left( \mathbb{R}_{+}\right) \right. \right\} $ of coherent
(exponential) functions $f^{\otimes }\left( \tau \right) =\bigotimes_{t\in
\tau }f^{\bullet }\left( t\right) $, given for each finite subset $\tau
=\left\{ t_1,...,t_n\right\} \subseteq \mathbb{R}_{+}$ by tensor products $%
f^{n_1,...,n_N}\left( \tau \right) =f^{n_1}\left( t_1\right)
...f^{n_N}\left( t_N\right) $, where $f^n,n=1,...,d$ are square-integrable
complex functions on $\mathbb{R}_{+}$ and $\xi ^f=0$ for almost all $%
f^{\bullet }=\left( f^n\right) $. The co-isometric shift $T_s$ intertwining $%
A^s\left( t\right) $ with $A\left( t\right) =T_sA^s\left( t\right)
T_s^{\dagger }$ is defined on $\mathcal{D}$ by $T_s\left( \eta \otimes
f^{\otimes }\right) \left( \tau \right) =\eta \otimes f^{\otimes }\left(
\tau +s\right) $. The complete positivity of the quantum stochastic adapted
map $\phi _t$ into the $\mathcal{D}$-forms $\left\langle \chi \right| \left.
\phi _t\left( X\right) \psi \right\rangle ,$ for $\chi ,\psi \in \mathcal{D}$
can be obviously written as 
\begin{equation*}
\sum_{k,l}\sum_{f,h}\left\langle \xi _k^f\right| \left. \phi _t\left(
f^{\bullet },X_{kl},h^{\bullet }\right) \xi _l^h\right\rangle \geq 0,\qquad
(5)
\end{equation*}
for any operator-matrix $\left[ X_{kl}\right] \geq 0$, where 
\begin{equation*}
\left\langle \eta \right| \left. \phi _t\left( f^{\bullet },X,h^{\bullet
}\right) \eta \right\rangle =\left\langle \eta \otimes f^{\otimes }\right|
\left. \phi _t\left( X\right) \eta \otimes h^{\otimes }\right\rangle
e^{-\int_t^\infty f^{\bullet }\left( s\right) ^{*}h^{\bullet }\left(
s\right) \mathrm{d}s},
\end{equation*}
$\xi ^f\neq 0$ only for a finite subset of $f^{\bullet }\in \left\{
f_i^{\bullet },i=1,2,...\right\} $. If the $\mathcal{D}$-form $\phi _t\left(
X\right) $ satisfies the stochastic equation (1), the $\mathfrak{h}$-form $%
\phi _t\left( f^{\bullet },X,h^{\bullet }\right) $ satisfies the
differential equation 
\begin{eqnarray*}
\frac{\mathrm{d}}{\mathrm{d}t}\phi _t\left( f^{\bullet },X,h^{\bullet
}\right) &=&f^{\bullet }\left( t\right) ^{*}h^{\bullet }\left( t\right) \phi
_t\left( f^{\bullet },X,h^{\bullet }\right) +\phi _t\left( f^{\bullet
},\alpha _{+}^{-}\left( X\right) ,h^{\bullet }\right) \\
&&+\sum_{m=1}^df^m\left( t\right) ^{*}\phi _t\left( f^{\bullet },\alpha
_{+}^m\left( X\right) ,h^{\bullet }\right) +\sum_{n=1}^dh^n\left( t\right)
\phi _t\left( f^{\bullet },\alpha _n^{-}\left( X\right) ,h^{\bullet }\right)
\\
&&+\sum_{m,n=1}^df^m\left( t\right) ^{*}h^n\left( t\right) \phi _t\left(
f^{\bullet },\alpha _n^m\left( X\right) ,h^{\bullet }\right)
\end{eqnarray*}
where $f^{\bullet }\left( t\right) ^{*}h^{\bullet }\left( t\right)
=\sum_{n=1}^df^n\left( t\right) ^{*}h^n\left( t\right) $. The positive
definiteness, (5), ensures the conditional positivity 
\begin{equation*}
\sum_{k,l}\sum_{f,h}\langle \xi _k^f|X_{kl}\xi _l^h\rangle =0\Rightarrow
\sum_{k,l}\sum_{f,h}\left\langle \xi _k^f\right| \left. \lambda _t\left(
f^{\bullet },X_{kl},h^{\bullet }\right) \xi _l^h\right\rangle \geq 0
\end{equation*}
of the form $\lambda _t\left( f^{\bullet },X,h^{\bullet }\right) =\frac
1t\left( \phi _t\left( f^{\bullet },X,h^{\bullet }\right) -X\right) $ for
each $t>0$ and any $\left[ X_{kl}\right] \geq 0.$ This applies also for the
limit $\lambda _0$ at $t\downarrow 0$, coinciding with the quadratic form 
\begin{equation*}
\left. \frac{\mathrm{d}}{\mathrm{d}t}\phi _t\left( f^{\bullet },X,h^{\bullet
}\right) \right| _{t=0}=\sum_{m,n}\bar{a}^m\lambda _n^m\left( X\right)
c^n+\sum_m\bar{a}^m\lambda ^m\left( X\right) +\sum_n\lambda _n\left(
X\right) c^n+\lambda \left( X\right) ,
\end{equation*}
where $a^{\bullet }=f^{\bullet }\left( 0\right) ,\quad c^{\bullet
}=h^{\bullet }\left( 0\right) $, and the $\lambda $'s are defined in (4).
Hence the form 
\begin{equation*}
\sum_{k,l}\sum_{\mu ,\nu }\left\langle \eta _k^\mu \right| \lambda _\nu ^\mu
\left( X_{kl}\right) \eta _l^\nu \rangle :=\sum_{k,l}\left\langle \eta
_k\right| \lambda \left( X_{kl}\right) \left| \eta _l\right\rangle
\end{equation*}
\begin{equation*}
+\sum_{k,l}\left( \sum_n\left\langle \eta _k\right| \left. \lambda _n\left(
X_{kl}\right) \eta _l^n\right\rangle +\sum_m\left\langle \eta _k^m\right|
\left. \lambda ^m\left( X_{kl}\right) \eta _l\right\rangle
+\sum_{m,n}\left\langle \eta _k^m\right| \left. \lambda _n^m\left(
X_{kl}\right) \eta _l^n\right\rangle \right)
\end{equation*}
with $\eta =\sum_f\xi ^f,\quad \eta ^{\bullet }=\sum_f\xi ^f\otimes
a^{\bullet }$, where $a^{\bullet }=f^{\bullet }\left( 0\right) $ is positive
if $\sum_{k,l}\langle \eta _k|X_{kl}\eta _l\rangle =0$ for a
positive-definite $\left[ X_{kl}\right] $. The components $\eta $ and $\eta
^{\bullet }$ of these vectors are independent because for any $\eta \in 
\mathfrak{h}$ and $\eta ^{\bullet }=\left( \eta ^1,...,\eta ^d\right) \in 
\mathfrak{h}\otimes \mathbb{C}^d$ there exists such a function $a^{\bullet
}\longmapsto \xi ^a$ on $\mathbb{C}^d$ with a finite support, that $%
\sum_a\xi ^a=\eta ,\quad \sum_a\xi ^a\otimes a^{\bullet }=\eta ^{\bullet },$
namely, $\xi ^a=0$ for all $a^{\bullet }\in \mathbb{C}^d$ except $a^{\bullet
}=0$, for which $\xi ^a=\eta -\sum_{n=1}^d\eta ^n$ and $a^{\bullet
}=e_n^{\bullet }$, the $n$-th basis element in $\mathbb{C}^d$, for which $%
\xi ^a=\eta ^n.$ This proves the complete positivity of the matrix form $%
\boldsymbol{\lambda }$, with respect to the matrix representation $%
\boldsymbol{\iota }$ defined in $\left( 4\right) $ on the ket-vectors $%
\boldsymbol{\eta }=\left( \eta ^\mu \right) $.%
\endproof%
\end{theorem}

\section{A Dilation Theorem for the Form-Generator.}

The conditional complete positivity of the structural map $\boldsymbol{%
\lambda }$ with respect to the degenerate representation $\boldsymbol{\iota }
$ written in the matrix form (4) obviously implies the positivity of the
dissipation form 
\begin{equation*}
\sum_{X,Z}\left\langle \boldsymbol{\eta }_X\right| \boldsymbol{\Delta }%
\left( X,Z\right) \left. \boldsymbol{\eta }_Z\right\rangle
:=\sum_{k,l}\sum_{\mu ,\nu }\left\langle \eta _k^\mu \right| \Delta _\nu
^\mu \left( X_k,X_l\right) \left. \eta _l^\nu \right\rangle ,\qquad \left(
6\right)
\end{equation*}
where $\eta ^{-}=\eta =\eta ^{+}$ and $\eta _k=\eta _{X_k}$ for any (finite)
sequence $X_k\in \mathcal{B},$ $k=1,2,...,$ corresponding to non-zero$%
\boldsymbol{\eta }_X\in \mathfrak{h}\otimes \mathbb{C}^{d+1}$. Here $%
\boldsymbol{\Delta }=\left( \Delta _\nu ^\mu \right) _{\nu \neq -}^{\mu \neq
+}$ is the stochastic dissipator, given by the blocks 
\begin{eqnarray*}
\Delta _n^m\left( X,Z\right) &=&\alpha _n^m\left( X^{\dagger }Z\right)
+X^{\dagger }Z\delta _n^m, \\
\Delta _n^{-}\left( X,Z\right) &=&\alpha _n^{-}\left( X^{\dagger }Z\right)
-X^{\dagger }\alpha _n^{-}\left( Z\right) =\Delta _{+}^n\left( Z,X\right)
^{\dagger } \\
\Delta _{+}^{-}\left( X,Z\right) &=&\alpha _{+}^{-}\left( X^{\dagger
}Z\right) -X^{\dagger }\alpha _{+}^{-}\left( Z\right) -\alpha _{+}^{-}\left(
X^{\dagger }\right) Z+X^{\dagger }DZ,
\end{eqnarray*}
where $D=\lambda \left( I\right) \leq 0$ ($D=0$ for the case of the
martingale $M_t$ ). In the linear case this means that the matrix-valued map 
$\lambda _{\bullet }^{\bullet }=\left[ \lambda _n^m\right] $, is completely
positive, and as follows from the next theorem, at least for the algebra $%
\mathcal{B}=\mathcal{B}\left( \mathfrak{h}\right) $ the maps $\lambda ,$ $%
\lambda ^m,$ $\lambda _n$ have the following form 
\begin{eqnarray*}
\lambda ^m\left( X\right) &=&\varphi ^m\left( X\right) -K_m^{\dagger
}X,\qquad \text{ }\lambda _n\left( X\right) =\varphi _n\left( X\right)
-XK_n\qquad \left( 7\right) \\
\lambda \left( X\right) &=&\varphi \left( X\right) -K^{\dagger }X-XK,\qquad 
\text{ }\varphi \left( I\right) \leq K+K^{\dagger }
\end{eqnarray*}
where $\boldsymbol{\varphi }=\left( \varphi _\nu ^\mu \right) _{\nu \neq
-}^{\mu \neq +}$ is a completely positive bounded map from $\mathcal{B}$
into the matrices of operators with the elements $\varphi _n^m=\lambda
_n^m,\varphi _{+}^m=\varphi ^m,\varphi _n^{-}=\varphi _n,\varphi
_{+}^{-}=\varphi :\mathcal{B}\rightarrow \mathcal{B}$.

In order to make the formulation of the dilation theorem as concise as
possible, we need the notion of the $\flat $-representation \cite{1} of the
unital $*$-multiplicative structure of the algebra $\mathcal{B}$ in a
pseudo-Hilbert space $\mathcal{E}=\mathfrak{h}\oplus \mathcal{K}\oplus 
\mathfrak{h}$ with respect to the indefinite metric 
\begin{equation*}
\left( \xi \right| \left. \xi \right) =2\func{Re}\left( \xi ^{-}\right|
\left. \xi ^{+}\right) +\left\| \xi ^{\circ }\right\| ^2+\left\| \xi
^{+}\right\| _D^2
\end{equation*}
for the triples $\xi =\left( \xi ^\mu \right) ^{\mu =-,\circ ,+}\in \mathcal{%
E}$, where $\xi ^{-},\xi ^{+}\in \mathfrak{h},\quad \xi ^{\circ }\in 
\mathcal{K},\quad \mathcal{K}$ is a pre-Hilbert space, and $\left\| \eta
\right\| _D^2=\left\langle \eta \right| \left. D\eta \right\rangle $. Define
the $\left( d+2\right) \times \left( d+2\right) $ matrix $\alpha =\left[
\alpha _\nu ^\mu \right] $ also for $\mu =+$ and $\nu =-,$ by 
\begin{equation*}
\alpha _\nu ^{+}\left( X\right) =0=\alpha _{-}^\mu \left( X\right) ,\qquad
\forall X\in \mathcal{B},
\end{equation*}
and then one can extend the summation in $\left( 1\right) $ so it is also
over $\mu =+$, and $\nu =-$. By such an extension the multiplication table
for $\mathrm{d}A\left( \alpha \right) =\alpha _\nu ^\mu \mathrm{d}A_\mu ^\nu 
$ can be written as 
\begin{equation*}
\mathrm{d}A\left( \beta \right) \mathrm{d}A\left( \gamma \right) =\mathrm{d}%
A\left( \beta \gamma \right)
\end{equation*}
in terms of the usual matrix product $\left( \beta \gamma \right) _\nu ^\mu
=\beta _\alpha ^\mu \gamma _\nu ^\alpha $ and the involution $\alpha \mapsto
\alpha ^{\flat }$ can be obtained by the pseudo-Hermitian conjugation $%
\alpha _\beta ^{\flat \nu }=G^{\nu \gamma }\alpha _\gamma ^{\mu *}G_{\mu
\beta }$ respectively to the indefinite metric tensor $G=\left[ G_{\mu \nu }%
\right] $ and its inverse $G^{-1}=\left[ G^{\mu \nu }\right] $, given by 
\begin{equation*}
G=\left[ 
\begin{array}{ccc}
0 & 0 & I \\ 
0 & I_{\circ }^{\circ } & 0 \\ 
I & 0 & D%
\end{array}
\right] ,\qquad G^{-1}=\left[ 
\begin{array}{ccc}
-D & 0 & I \\ 
0 & I_{\circ }^{\circ } & 0 \\ 
I & 0 & 0%
\end{array}
\right]
\end{equation*}
with an arbitrary $D$, where $I_{\circ }^{\circ }$ is the identity operator
in $\mathcal{K}$, being equal $I_{\bullet }^{\bullet }=\left[ I\delta _n^m%
\right] _{n=1,...,d}^{m=1,...,d}$ in the case of $\mathcal{K}=\mathfrak{h}%
\otimes \mathbb{C}^d.$

\begin{theorem}
The following are equivalent:

\begin{enumerate}
\item The dissipation form (6), defined by the $\flat $-map $\alpha $ with $%
\alpha _{+}^{-}\left( I\right) =D$, is positive definite: $%
\sum_{X,Z}\left\langle \boldsymbol{\eta }_X\right| \boldsymbol{\Delta }%
\left( X,Z\right) \left. \boldsymbol{\eta }_Z\right\rangle \geq 0.$

\item There exists a pre-Hilbert space $\mathcal{K}$, a unital $*$-
representation $j$ in $\mathcal{B}\left( \mathcal{K}\right) ,$%
\begin{equation*}
j\left( X^{\dagger }Z\right) =j\left( X\right) ^{\dagger }j\left( Z\right)
,\quad j\left( I\right) =I,
\end{equation*}
of the multiplication structure of $\mathcal{B}$, a $\left( j,i\right) $%
-derivation of $\mathcal{B}$ with $i\left( X\right) =X$, 
\begin{equation*}
k\left( X^{\dagger }Z\right) =j\left( X\right) ^{\dagger }k\left( Z\right)
+k\left( X^{\dagger }\right) Z,
\end{equation*}
having values in the operators $\mathfrak{h}\rightarrow \mathcal{K}$, the
adjoint map $k^{*}\left( X\right) =k\left( X^{\dagger }\right) ^{\dagger }$,
with the property 
\begin{equation*}
k^{*}\left( X^{\dagger }Z\right) =X^{\dagger }k^{*}\left( Z\right)
+k^{*}\left( X^{\dagger }\right) j\left( Z\right)
\end{equation*}
of $\left( i,j\right) $-derivation in the operators $\mathcal{K}\rightarrow 
\mathfrak{h}$, and a map $l:\mathcal{B}\rightarrow \mathcal{B}$ having 
\begin{equation*}
l\left( X^{\dagger }Z\right) =X^{\dagger }l\left( Z\right) +l\left(
X^{\dagger }\right) Z+k^{*}\left( X^{\dagger }\right) k\left( Z\right) ,
\end{equation*}
with the adjoint $l^{*}\left( X\right) =l\left( X\right) +\left[ D,X\right]
,\quad $such that $\lambda \left( X\right) =l\left( X\right) +DX,$%
\begin{equation*}
\lambda _n\left( X^{\dagger }\right) =k\left( X\right) ^{\dagger }L_n^{\circ
}+X^{\dagger }L_n^{-}=\lambda ^n\left( X\right) ^{\dagger },
\end{equation*}
and $\lambda _n^m\left( X\right) =L_m^{\circ \dagger }j\left( X\right)
L_n^{\circ }$ for some operators $L_n^{\circ }:\mathfrak{h}\rightarrow 
\mathcal{K}$ having the adjoints $L_n^{\circ \dagger }$ on $\mathcal{K}$ and 
$L_n^{-}\in \mathcal{B}$.

\item There exists a pseudo-Hilbert space, $\mathcal{E}$, namely, $\mathfrak{%
h}\oplus \mathcal{K}\oplus \mathfrak{h}$ with the indefinite metric tensor $%
G=\left[ G_{\mu \nu }\right] $ given above for $\mu ,\nu =-,\circ ,+$, and $%
D=\lambda \left( I\right) ,$ a unital $\flat $-representation $\jmath =\left[
\jmath _\nu ^\mu \right] _{\nu =-,\circ ,+}^{\mu =-,\circ ,+}$ of the
multiplication structure of $\mathcal{B}$ on $\mathcal{E}$ : 
\begin{equation*}
\jmath \left( X^{\dagger }Z\right) =\jmath \left( X\right) ^{\flat }\jmath
\left( Z\right) ,\quad \jmath \left( I\right) =I
\end{equation*}
with $\jmath \left( X\right) ^{\flat }=G^{-1}\jmath \left( X\right)
^{\dagger }G$, given by 
\begin{equation*}
\jmath _{\circ }^{\circ }=j,\quad \jmath _{+}^{\circ }=k,\quad \jmath
_{\circ }^{-}=k^{*},\quad \jmath _{+}^{-}=l,\quad \jmath _{-}^{-}=i=\jmath
_{+}^{+}
\end{equation*}
and all other $\jmath _\nu ^\mu =0,$ and a linear operator $\mathbf{L}:%
\mathfrak{h}\oplus \mathfrak{h}^{\bullet }\rightarrow \mathcal{E}$, where $%
\mathfrak{h}^{\bullet }=\mathfrak{h}\otimes \mathbb{C}^d$, with the
components $\left( L^\mu ,L_{\bullet }^\mu \right) ,$%
\begin{equation*}
L^{-}=0,\quad L^{\circ }=0,\quad L^{+}=I,\quad L_{\bullet }^{-}=\left(
L_n^{-}\right) ,\quad L_{\bullet }^{\circ }=\left( L_n^{\circ }\right)
,\quad L_{\bullet }^{+}=0,
\end{equation*}
and $\mathbf{L}^{\flat }=\left( 
\begin{array}{ccc}
I & 0 & D \\ 
0 & L_{\circ }^{\bullet } & L_{+}^{\bullet }%
\end{array}
\right) =\mathbf{L}^{\dagger }G$, where $L_{\circ }^{\bullet }=L_{\bullet
}^{\circ \dagger },L_{+}^{\bullet }=L_{\bullet }^{-\dagger }$ , such that 
\begin{equation*}
\mathbf{L}^{\flat }\jmath \left( X\right) \mathbf{L}=\boldsymbol{\lambda }%
\left( X\right) ,\qquad \forall X\in \mathcal{B}.
\end{equation*}

\item The structural map $\boldsymbol{\lambda }=\boldsymbol{\alpha }+%
\boldsymbol{\delta }$ is conditionally positive-definite with respect to the
matrix representation $\boldsymbol{\iota }$ in $\left( 4\right) $ : 
\begin{equation*}
\sum_X\boldsymbol{\iota }\left( X\right) \boldsymbol{\eta }%
_X=0\Longrightarrow \sum_{X,Z}\langle \boldsymbol{\eta }_X|\boldsymbol{%
\lambda }\left( X^{\dagger }Z\right) \boldsymbol{\eta }_Z\rangle \geq 0.
\end{equation*}
\end{enumerate}
\end{theorem}

\proof
Similar to the dilation theorem in \cite{4}, see also \cite{5}.%
\endproof%

\section{The structure of the w*-analytical CP cocycles with bounded
generators.}

The structure $\left( 7\right) $ of the linear form- generator for CP
cocycles over $\mathcal{B}=\mathcal{B}\left( \mathfrak{h}\right) $ is a
consequence of the well known fact that the linear derivations $k,k^{*}$ of
the algebra $\mathcal{B}\left( \mathfrak{h}\right) $ of all bounded
operators on a Hilbert space $\mathfrak{h}$ are spatial, $k\left( X\right)
=j\left( X\right) L-LX,\quad k^{*}\left( X\right) =L^{\dagger }j\left(
X\right) -XL^{\dagger }, $ and so 
\begin{equation*}
l\left( X\right) =\frac 12\left( L^{\dagger }k\left( X\right) +k^{*}\left(
X\right) L+\left[ X,D\right] \right) +i\left[ H,X\right] ,
\end{equation*}
where $H^{\dagger }=H$ is a Hermitian operator in $\mathfrak{h}$. The
structural map $\boldsymbol{\lambda }$ whose components are composed (as in $%
\left( 7\right) $) into the sums of the components $\varphi _\nu ^\mu \quad $%
of a CP map $\boldsymbol{\varphi }:\mathcal{B}\rightarrow \mathcal{B}\otimes 
\mathcal{M}\left( \mathbb{C}^{d+1}\right) $ and left and right
multiplications, are obviously conditionally completely positive with
respect to the representation $\boldsymbol{\iota }$ in (4). As follows from
the dilation theorem in this case, there exists a family $%
L_{-}=L=L_{+},\quad L_n=L_n^{\circ },\quad n=1,...,d$ of linear operators $%
L_\nu :\mathfrak{h}\rightarrow \mathcal{K},$ having adjoints $L_\mu
^{\dagger }:\mathcal{K}\rightarrow \mathfrak{h}$ such that $\varphi _\nu
^\mu \left( X\right) =L_\mu ^{\dagger }j\left( X\right) L_\nu ,.$

The next theorem proves that these structural conditions which are
sufficient for complete positivity of the cocycles, given by the equation
(1), are also necessary if the structural map $\boldsymbol{\lambda }$ is
w*-analytic \cite{6} and bounded on the unit ball of a von-Neumann algebra $%
\mathcal{B}$. Thus the equation $\left( 1\right) $ for a completely positive
w*-analytical cocycle with bounded stochastic derivatives has the following
general form 
\begin{equation*}
\mathrm{d}\phi _t\left( X\right) +\phi _t\left( K^{\dagger }X+XK-L^{\dagger
}j\left( X\right) L\right) \mathrm{d}t=\sum_{m,n=1}^d\phi _t\left(
L_m^{*}j\left( X\right) L_n-X\delta _n^m\right) \mathrm{d}A_m^n
\end{equation*}
\begin{equation*}
+\sum_{m=1}^d\phi _t\left( L_m^{*}j\left( X\right) L-K_m^{*}X\right) \mathrm{%
d}A_m^{+}+\sum_{n=1}^d\phi _t\left( L^{*}j\left( X\right) L_n-XK_n\right) 
\mathrm{d}A_{-}^n,\qquad (8)
\end{equation*}
given by a w*-analytical representation $j$. This gives a quantum stochastic
generalization of the branching norm-continuous semigroups with the
nonlinear generators \cite{6} . If the space $\mathcal{K}$ can be embedded
into the direct sum $\mathfrak{h}\otimes \mathbb{C}^d=\mathfrak{h}\oplus
...\oplus \mathfrak{h}$ of $d$ copies of the initial Hilbert space $%
\mathfrak{h}$ such that $j\left( X\right) =\left[ X\delta _n^m\right] $,
this equation corresponds to the Lindblad form \cite{7} for the generator $%
\lambda =\alpha _{+}^{-}$. But in the contrast to the Lindblad equation, it
can be resolved in the form $\phi _t\left( X\right) =F_t^{\dagger }XF_t$,
where $F=\left( F_t\right) _{t>0}$ is an (unbounded) cocycle in the tensor
product $\mathfrak{h}\otimes \mathcal{F}$ with Fock space $\mathcal{F}$ over
the Hilbert space $\mathbb{C}^d\otimes L^2\left( \mathbb{R}_{+}\right) $ of
the quantum noise of dimensionallity $d$. The cocycle $F$ satisfies the
quantum stochastic equation 
\begin{equation*}
\mathrm{d}F_t+KF_t\mathrm{d}t=\sum_{i,n=1}^d\left( L_n^i-I\delta _n^i\right)
F_t\mathrm{d}A_i^n+\sum_{i=1}^dL^iF_t\mathrm{d}A_i^{+}-\sum_{n=1}^dK_nF_t%
\mathrm{d}A_{-}^n,\qquad \left( 9\right)
\end{equation*}
where $L_n^{i\text{ }}$ and $L^i$ are the operators in $\mathfrak{h}$,
defining 
\begin{eqnarray*}
\varphi _n^m\left( X\right) &=&\sum_{i=1}^dL_m^{i\dagger }XL_n^i,\qquad
\varphi \left( X\right) =\sum_{i=1}^dL^{i\dagger }XL^i \\
\varphi ^m\left( X\right) &=&\sum_{i=1}^dL_m^{i\dagger }XL^i,\qquad \varphi
_n\left( X\right) =\sum_{i=1}^dL^{i\dagger }XL_n^i\qquad \left( 10\right)
\end{eqnarray*}
with $\sum_{i=1}^dL^{i\dagger }L^i=K+K^{\dagger }$if $M_t$ is a martingale ($%
\leq K+K^{\dagger }$if submartingale) .\allowbreak

\begin{theorem}
Let the structural maps $\boldsymbol{\lambda }$ of the quantum stochastic
cocycle $\phi $ over a von-Neumann algebra $\mathcal{B}$ be w*-analytic and
bounded: 
\begin{equation*}
\left\| \lambda \right\| <\infty ,\qquad \left\| \lambda _{\bullet }\right\|
=\left( \sum_{n=1}^d\left\| \lambda _n\right\| ^2\right) ^{\frac 12}=\left\|
\lambda ^{\bullet }\right\| <\infty ,\qquad \left\| \lambda _{\bullet
}^{\bullet }\right\| =\left\| \lambda _{\bullet }^{\bullet }\left( I\right)
\right\| <\infty ,
\end{equation*}
where $\left\| \lambda \right\| =\sup \left\{ \left\| \lambda \left(
X\right) \right\| :\left\| X\right\| <1\right\} ,\quad \left\| \lambda
_{\bullet }^{\bullet }\left( I\right) \right\| =\sup \left\{ \left\langle
\eta ^{\bullet },\lambda _{\bullet }^{\bullet }\left( I\right) \eta
^{\bullet }\right\rangle \left| \left\| \eta ^{\bullet }\right\| <1\right.
\right\} ,$ and $\phi _t$ is a CP cocycle, satisfying equation (1) with $%
\phi _0\left( X\right) =X.$ Then they have the form (7) written as 
\begin{equation*}
\boldsymbol{\lambda }\left( X\right) =\boldsymbol{\varphi }\left( X\right) -%
\boldsymbol{\iota }\left( X\right) \boldsymbol{K}-\boldsymbol{K}^{\dagger }%
\boldsymbol{\iota }\left( X\right)
\end{equation*}
with $\varphi =\varphi _{+}^{-},\quad \varphi ^m=\varphi _{+}^m,\quad
\varphi _n=\varphi _n^{-}$ and $\varphi _n^m=\lambda _n^m$, composing a
w*-analytical bounded CP map. 
\begin{equation*}
\boldsymbol{\varphi }=\left( 
\begin{array}{cc}
\varphi & \varphi _{\bullet } \\ 
\varphi ^{\bullet } & \varphi _{\bullet }^{\bullet }%
\end{array}
\right) ,\quad and\quad \boldsymbol{K}=\left( 
\begin{array}{cc}
K & K_{\bullet } \\ 
K^{\bullet } & K_{\bullet }^{\bullet }%
\end{array}
\right)
\end{equation*}
with arbitrary $K^{\bullet },K_{\bullet }^{\bullet }$. The equation (8) has
the unique CP solution, given by the iteration of the quantum stochastic
integral equation 
\begin{equation*}
\phi _t\left( X\right) =V_t^{\dagger }XV_t+\int_0^tV_s^{\dagger }\beta _\nu
^\mu \left( \phi _{t-s}^s\left( X\right) \right) V_s\mathrm{d}A_\mu ^\nu
\left( s\right)
\end{equation*}
where $\beta _\nu ^\mu \left( X\right) =\varphi _\nu ^\mu \left( X\right)
-X\delta _\nu ^\mu $ and $V_t$ is the vector cocycle $V_r^sV_s=V_{r+s}$,
resolving the quantum stochastic differential equation 
\begin{equation*}
\mathrm{d}V_t+KV_t\mathrm{d}t+\sum_{m=1}^dK_mV_t\mathrm{d}A_{-}^m=0
\end{equation*}
with the initial condition $V_0=I$ in $\mathfrak{h}$ and with $%
V_r^s=T_r^{\dagger }V_rT_s$, shifted by the time-shift co-isometry $T_s$ in $%
\mathcal{D}$.
\end{theorem}

\end{document}